\documentclass[preprint,aps,nofootinbib]{revtex4}
\usepackage{graphicx}
\usepackage{epstopdf}
\usepackage{subfigure}
\usepackage{textcomp}
\usepackage{dsfont}
\usepackage{amsfonts}
\usepackage{amsmath}
\usepackage{amssymb}
\def\bkR{{\rm I\kern-.17em R}}
\def\bkC{{\rm \kern.24em \vrule width.05em height1.4ex depth-.05ex \kern-.26em C}}

\def\be{\beta}

\def\Pb{\overline{\Psi}}

\def\frac#1#2{{\textstyle{{#1}\over {#2}}}}

\def\lsim{\mathrel{\rlap{\lower4pt\hbox{\hskip1pt$\sim$}}
    \raise1pt\hbox{$<$}}}
\def\gsim{\mathrel{\rlap{\lower4pt\hbox{\hskip1pt$\sim$}}
    \raise1pt\hbox{$>$}}}
\def\sqr#1#2{{\vcenter{\vbox{\hrule height.#2pt
         \hbox{\vrule width.#2pt height#1pt \kern#1pt
         \vrule width.#2pt}
         \hrule height.#2pt}}}}

\def\laq{\raise 0.4 ex \hbox{$<$}\kern -0.8 em\lower 0.62 ex\hbox{$\sim$}}
\def\gaq{\raise 0.4 ex \hbox{$>$}\kern -0.7 em\lower 0.62 ex\hbox{$\sim$}}

\def\be{\begin{equation}}
\def\ee{\end{equation}}
\def\beqa{\begin{eqnarray}}
\def\eeqa{\end{eqnarray}}

\def\dalemb#1#2{{\vbox{\hrule height.#2pt
        \hbox{\vrule width.#2pt height#1pt \kern#1pt \vrule width.#2pt}
        \hrule height.#2pt}}}

\def\dalemb#1#2{{\vbox{\hrule height.#2pt
        \hbox{\vrule width.#2pt height#1pt \kern#1pt \vrule width.#2pt}
        \hrule height.#2pt}}}

\def\gtorder{\mathrel{\raise.3ex\hbox{$>$}\mkern-14mu
             \lower0.6ex\hbox{$\sim$}}}
\def\ltorder{\mathrel{\raise.3ex\hbox{$<$}\mkern-14mu
             \lower0.6ex\hbox{$\sim$}}}

\begin{document}

\rightline{February 2015}

\title{A Phase-Space Noncommutative Picture of Nuclear Matter}

\author{Orfeu Bertolami\footnote{E-mail: orfeu.bertolami@fc.up.pt}}

\author{Hodjat Mariji\footnote{E-mail: hmariji@ut.ac.ir; astrohodjat@fc.up.pt}}

\vskip 0.3cm

\affiliation{Departamento de F\'\i sica e Astronomia, Faculdade de Ci\^encias da Universidade do Porto \\
Rua do Campo Alegre 687, 4169-007 Porto, Portugal}

\vskip 0.5cm

\begin{abstract}

\vskip 0.5cm

{
Noncommutative features are introduced into a relativistic quantum field theory model of nuclear matter, the quantum hadrodynamics-I nuclear model (QHD-I). It is shown that the nuclear matter equation of state (NMEoS) depends on the fundamental momentum scale, $\eta$, introduced by the phase-space noncommutativity (NC). Although it is found that NC geometry does not affect the nucleon fields up to $O(\eta^2)$, it affects the energy density, the pressure and other derivable quantities of the NMEoS, such as the nucleon \textit{effective mass}. Under the conditions of saturation of the symmetric NM, the estimated value for the noncommutative parameter is $\sqrt{\eta}=0.014 MeV/c$.  
}

\end{abstract}

\maketitle

\section{Introduction}

Introducing noncommutative geometric features is believed to be an interesting way to generalize quantum mechanics \cite{1, 2, 3, 4, 5, 6}. Noncommutative quantum mechanics (NCQM) arises as deformations of the Heisenberg-Weyl algebra. NCQM lives in a 2\textit{d}-dimensional phase-space, where time is assumed to be a commutative parameter, and coordinate and momentum variables obey a NC algebra \cite{5, 6}:
\begin{align}
[x^i, x^j]=i\theta\epsilon^{ij},&& [p^i,p^j]=i\eta\epsilon^{ij},&& [x^i,p^j]=i\hbar_{eff}\delta^{ij},\label{first}
\end{align}
where $\epsilon_{ij}$ is an antisymmetric matrix \footnote{It is used that ${1\over 2}\varepsilon_{ijk}\epsilon_{ij}=e_k$, $\varepsilon_{ijk}\epsilon_{ij}$ and $e_k$ are Levi-Civita tensors and a normalized vector, respectively.}, \textit{i} and \textit{j} range from 1 to 3, and 
\begin{align}
\hbar_{eff}=\hbar\left( 1 + {\theta\eta \over 4\hbar^2} \right) .
\end{align}

The NC parameters, $\theta$ and $\eta$, are believed to be new fundamental constants of nature together with Planck's constant. The fundamental scales of NC geometry are $l_{\theta}=\sqrt{\theta}$ and $l_{\eta}=\sqrt{\eta}$ which must be obtained from experimental data for specific systems; for instance, the following bounds have been set \cite{5, 7, 8}:
\begin{align}
l_{\theta} \le 2\times10^{-5}  fm ,&& l_{\eta} \le 8\times10^{-1}   meV/c.
\end{align}

The NC extensions of QM show an impressive range of implications: on the quantum Hall effect \cite{8}, on the Landau level and the 2\textit{D} harmonic oscillator problems in the phase-space \cite{9, 10}, and as a probe for quantum beating and missing information effects \cite{11} and as a source for quantum entanglement \cite{12}. NCQM also admits violations of the Roberston-Schr\"odinger and Ozawa's uncertainty relations \cite{13}. Furthermore, in the framework of quantum cosmology, phase-space noncommutativity has shown to give origin to the novel features for the black hole singularity \cite{14, 15, 16, 17}, as well as to the equivalence principle \cite{18}. One also expects some implications to compact objects \cite{19}. 

In the present work, we examine the impact of NC in nuclear physics, in particular, on relativistic nuclear matter (NM) calculations. Nowadays, there is a growing interest in applications of the primary NM theories \cite{20,21,22,23,24,25} to study compact stellar objects. Of course, this has bearings on the mass-radius relationship and on studies of the crust thickness of neutron stars (NS), and on conditions for the collapse of NS and black holes. Other implications include supernovae explosions, the study of energetic heavy ions collisions \cite{26}, and on the properties of ordinary nuclei \cite{27}.
 
   In order to examine the NC effects into the NM calculations, we consider the QHD-I (or the $\sigma\omega$-) model, a well known renormalizable relativistic quantum field model of the nucleon (p, n) system, interacting with the neutral scalar and vector mesons, $\sigma$, and, $\omega$. Moreover, following Ref. \cite{28}, we assume that at high baryon densities, the scalar and the vector fields are replaced by their expectation values, which serve as a mean field in which the nucleons move. As will be shown, for a suitable value of $\eta$ under the empirical saturation conditions, the \textit{effective mass} of a Dirac nucleon, $ M^{*}$ (cf. Eqs. (20) and (37)), has a reasonable value.
   
   This paper is organized as follows. In section II, we briefly set the noncommutative tools to be used in our nuclear model. The field approach required to tackle the nuclear problem, the QHD-I model, is discussed in section III. In section IV we consider the NC QHD-I model. Finally, in section V, we present our results and discussion.

\section{Some NC Geometry Tools}

In order to generalize the QHD-I model subject to the algebra Eqs. (1), we must consider (see e.g., Ref. \citep{29}) that NC fields satisfy the generalized Moyal $\star$ product \cite{6}:
\begin{align}
f(x)\star g(x)\equiv f(x)\exp\left\{{i\over 2}\overleftarrow{\partial_i}\theta\overrightarrow{\partial_j}\right\}g(x)\approx f(x)g(x)+{i\over 2}\theta_{ij}\partial_if(x)\partial_jg(x)+\mathcal{O}(\theta^2).\label{moyal}
\end{align}
This truncation will be sufficient for the purposes of our study and the dependence on the $\eta$ parameter will arise through the so called Seiberg-Witten map \cite{30}.
This is a noncanonical transformation, $(x,p)\mapsto(x',p')$, that maps the NC algebra into the Heisenberg-Weyl algebra \cite{6}:
\begin{align}
[x'_i,x'_j]=0, && [p'_i,p'_j]=0, && [x'_i,p'_j]=i\hbar\delta_{ij}.
\end{align}
The NC variables, Eq. (1), can be mapped into the commutative ones, Eq. (5), through the Seiberg-Witten map:
\begin{align}
x_i=ax'_i-\left({\theta \over 2a\hbar}\right)\epsilon_{ij}p'_j, && p_i=bp'_i+\left({\eta\over 2b\hbar}\right)\epsilon_{ij}x'_j,\label{map}
\end{align}
where without loss of generality we choose $\textit{a=b=1}$ \cite{6, 31}.

In what follows we adapt the phase-space NC treatment of Dirac fields developed in Ref. \citep{29} to the QHD-I model. As will seen this will allows us to directly assess the effect of the dependence on the NC parameter on the well known NM quantities.

\section{A Brief Review of the QHD-I Model}

   We briefly review here the QHD-I ($\sigma\omega$-) model (see also Ref. \cite{28}). In the QHD-I model, the neutral scalar meson field, $\Phi$, couples to the scalar density of baryons, $\Pb\Psi$, through the Yukawa interaction term $g_{s}\Pb\Psi\Phi$ with strength given by the coupling constant $g_{s}$, while the neutral vector meson, $V^{\mu}$, couples to the conserved baryon current, $\Pb\gamma^{\mu}\Psi$, through $g_{v}\Pb\gamma_{\mu}\Psi V^{\mu}$ with the coupling constant $g_{v}$. In the mean field approach (MFA), baryons are assumed to move in a box of volume $\Omega$ within the mean field of the expectation values of the constant and condensed scalar and vector fields, $\Phi_{0}$ and $V_{0}$, respectively. The \textit{effective mass} of a Dirac nucleon, $M^{*}$, is given by $M^{*}=M-g_{s}\Phi_{0}$, and the Lagrangian density for QHD-I model, in the MFA, is as follows \cite{28}:
\begin{align}
\mathcal{L}_{QHD-I}=\Pb\left(i\gamma^{\mu}\partial_{\mu}-\beta g_{v}V_{0}-M^{*}\right)\Psi+C_{0},\label{lagrngC} 
\end{align}
where $C_{0}$ is written via the constant scalar and vector meson mean fields \cite{28}
\begin{align}
C_{0}={1 \over 2 } m_{s}^{2}\Phi_{0}^{2}-{1 \over 2 }m_{v}^{2}V_{0}^{2}.
\end{align}
In Eq. (7), $\gamma_{\mu}=\left(\gamma_{0},-\gamma_{i}\right) $, and 
\begin{align}
\gamma_{0}=\beta,&&\gamma_{i}=\beta\alpha_{i}.
\end{align}   		
In Eqs. (9), the Dirac matrices, $\beta$ and $\alpha_{i}$, satisfy the following relations:
\begin{align}
[\alpha_i,\alpha_j]_{+}=2\delta_{ij}, && [\alpha_i,\beta]_{+}=0, && \alpha_i^2=\beta^2=\mathds{1},
\end{align}
so that 
\begin{equation}
\alpha_i=
\begin{bmatrix}
0 & \sigma_i \\
\sigma_i & 0
\end{bmatrix}, \hspace{30pt}
\beta=
\begin{bmatrix}
\mathds{1} & 0 \\
0 & \mathds{1}
\end{bmatrix} ,
\end{equation}
where $\mathds{1}$ and $\sigma_{i}$ are the well known $2\times2$ unit and Pauli matrices, respectively. Using Eqs. (7) and (8), through the field equations, we obtain the following relations and the equation of motion for $\Psi$:
\begin{align}
\Phi_{0}={g_{s} \over m_{s}^{2} }\rho_{s},\\
V_{0}={g_{v} \over m_{v}^{2} }\rho_{B},\\
\left(i\gamma^{\mu}\partial_{\mu}-\beta g_{v}V_{0}-M^{*}\right)\Psi=0,
\end{align}
where $\rho_{s}=\left\langle \Pb\Psi\right\rangle  $ and $\rho_{B}=\left\langle \Psi^{\dagger}\Psi\right\rangle $ are the scalar and baryon density, respectively, and the brackets $< >$ denote the vacuum expectation values. Since the fields are assumed to be constant, there is a static and uniform set of particles. Thus, considering free nucleons $\psi(\textbf{k},\lambda)\lbrace exp\left[i\textbf{k.x}-i\varepsilon(k)t\right] \rbrace$ with momentum $\textbf{k}$, energy $ \varepsilon(k)=\varepsilon(|\textbf{k}|)$, polarization $\lambda$, and a four-component Dirac spinor, $\psi(\textbf{k},\lambda)$, Eq. (14) leads to
\begin{align}
\left[\boldsymbol{\alpha}.\textbf{k}+\beta M^{*} \right]\psi\left( \textbf{k},\lambda\right) =\varepsilon^{*}\left( k\right) \psi\left( \textbf{k},\lambda\right),
\end{align}
where $\varepsilon^{*}(k)=\left[\varepsilon(k)-g_{v}V_{0}\right] $ is the $\textit{effective energy}$ of a nucleon. Regarding the superposition of solutions of Eq. (14), its general solution is given by:

\begin{equation}
\Psi(\textbf{\textit{x}},\textit{t})={1 \over \sqrt{\Omega} }\sum_{\textbf{k}, \lambda} \left\lbrace  A_{\textbf{k}\lambda}U(\textbf{k},\lambda) e^{i[\textbf{k.x}-\varepsilon_{+}(k)t]}+B_{\textbf{k}\lambda}^{\dagger}V(\textbf{k},\lambda)e^{-i[\textbf{k.x}+\varepsilon_{-}(k)t]} \right\rbrace ,
\end{equation}
where $U(V)$ is the positive (negative)-energy spinor. A straightforward calculation, shows that:
\begin{align}
\varepsilon_{\pm}(k)=\left[g_{v}V_{0} \pm E^{*}(k)\right],
\end{align}
where $ E^{*}(k)=\sqrt{\textbf{k}^{2}+M^{*2}} $. It should be pointed out that in Eq. (16), the summation over $|\textbf{k}|$ is limited to $k_{F}$, the Fermi momentum of the nucleons. On the other hand, the summation over $\lambda$ comprises summation over spin and iso-spin.
Finally, through the familiar relationship between the energy-momentum tensor, energy density and pressure, one can write the NMEoS in the QHD-I framework \cite{28}:
\begin{equation}
\varepsilon_{C}={1 \over 2}\left( { g_{v} \over m_{v}}\rho_{B}\right)^{2} + {1 \over 2}{\left[ {m_{s}\over g_{s}}(M-M^{*})\right]^{2}} + {\nu \over 2\pi^{2}}\int_{0}^{k_{F}} E^{*}(k)k^{2}dk, 
\end{equation}
\begin{equation}
p_{C}={1 \over 2}\left( { g_{v} \over m_{v}}\rho_{B}\right)^{2} - {1 \over 2}{\left[ {m_{s}\over g_{s}}(M-M^{*})\right]^{2}} +{1 \over 3}\left( {\nu \over 2\pi^{2}}\right) \int_{0}^{k_{F}} {k^{4}\over E^{*}(k)}dk, 
\end{equation}
where $\nu$, the degeneracy on the spin and iso-spin of nucleons, is 4. Of course, Eq. (19) follows from Eq. (18) through the relationship $ p=\rho_{B}^{2} \left[ \partial\left( \varepsilon/\rho_{B} \right)/ \partial\rho_{B} \right] $. The energy of an isolated system is made minimal at fixed volume, $\Omega$, baryon number, $B$, and temperature (here vanishing), by minimizing $\varepsilon$ with respect to $M^{*}$. Now, $M^{*}$ is given by the self-consistency relation:
\begin{align}
M^{*}=M-{\left({g_{s}\over m_{s}}\right)^{2}}{\nu \over 2\pi^{2}}\int_{0}^{k_{F}} k^{2}{M^{*}\over E^{*}(k)}dk.
\end{align}

\section{The NC QHD-I Model}

The NMEoS, Eqs. (18) and (19), as well as the nucleon \textit{effective mass}, Eq. (20), are obtained from the Lagrangian density, Eq. (7), in the MFA approach. These relations are changed by the NC algebra, Eqs. (1). Following Ref. \cite{29}, the ordinary product of fields in the Lagrange density is replaced by the Moyal product, $\star$, for the NC fields, $\Psi'$. In the MFA, the Lagrangian density, in the NC geometry is as follows:
\begin{align}
\mathcal{L'}_{QHD-I}=\Pb '\star\left(i\gamma^{\mu}\partial_{\mu}-\beta g_{v}V_{0}-M^{*}\right)\star\Psi'+C_{0}.\label{lagrngNC} 
\end{align}
As described in Section II, the NC fields and variables must be mapped into commutative ones. Thus, using the transformations Eq. (6), we get for the free noncommutative Dirac fields:
\begin{align}
\Psi'\left(\textbf{x}',t\right) \sim e^{i\textbf{k}'.\textbf{x}'}, 
\end{align}
where it has been used that of $k'_{i}=p'_{i}=-i\partial'_{i}$. Hence the NC Lagrangian density of QHD-I reads:
\begin{align}
\mathcal{L'}_{NC}&=\Pb'(i\gamma^{\mu} \partial'_{\mu}-{1 \over 2} \eta_{kl} \gamma^{k} x'^{l}-\beta g_{v}V_{0}-M^{*} )\Psi' + C_{0},
\end{align}
where $i\gamma^{i}\partial_{i}=i\gamma^{i} \partial'_{i}-{1 \over 2} \eta_{kl} \gamma^{k} x'^{l}$, and $\theta=0$ was set in order to preserve gauge invariance (see Ref. \cite{29}).
Therefore, the equation of motion for the Dirac field reads
\begin{align}
\left(i\gamma^{\mu}\partial'_{\mu}-{1 \over 2} \eta_{kl} \gamma^{k} x'^{l}-\beta g_{v}V_{0}-M^{*}\right)\Psi'=0.
\end{align}
Eq. (15) is changed to
\begin{align}
\left[\boldsymbol{\alpha}.\textbf{k}'+ {1 \over 2} \boldsymbol{\alpha} \times \textbf{x}'.\boldsymbol{\eta}+\beta M^{*} \right]\psi'\left( \textbf{k}',\lambda \right)= \varepsilon'^{*}\left( k'\right) \psi'\left( \textbf{k}',\lambda\right),
\end{align}
where $\varepsilon'^{*}=\varepsilon'-g_{v}V_{0}$. For the spectral energy of the positive and negative states, $\varepsilon'_{\pm}$, one finds:
\begin{align}
\varepsilon'_{\pm} = \left[ g_{v} V_{0} \pm E^{'*}_{\eta} \right] ,
\end{align}
with $ E^{'*}_{\eta}=\sqrt{\textbf{k}'^{2}+M^{*2}+\textbf{k}'\times\textbf{x}'.\boldsymbol{\eta}} $. For the energy density  one finds:
\begin{equation}
\varepsilon'= {1 \over 2} \left( { g_{v} \over m_{v}}\rho_{B}\right)^{2} + {1 \over 2}{\left[ {m_{s}\over g_{s}}(M-M^{*})\right]^{2}} + {\nu \over \Omega^{2}}\sum_{|\textbf{k}'|\leq k_{F}} \int E^{'*}_{\eta}d\textbf{x}'. 
\end{equation}
In a similar way, one can get for the pressure:
\begin{equation}
p'={1 \over 2}\left( { g_{v} \over m_{v}}\rho_{B}\right)^{2} - {1 \over 2}{\left[ {m_{s}\over g_{s}}(M-M^{*})\right]^{2}} +{1 \over 3}\left( {\nu \over \Omega^{2}}\right) \sum_{|\textbf{k}'|\leq k_{F}} \int {\textbf{k}'^{2}\over E^{'*}_{\eta}}d\textbf{x}', 
\end{equation}

Expanding $E^{'*}_{\eta}$ up to $O(\eta^{2})$, and converting the sum over $\textbf{k}'$ into an integral, hence:
\begin{equation}
\varepsilon'={1 \over 2}\left( { g_{v} \over m_{v}}\rho_{B}\right)^{2} + {1 \over 2}{\left[ {m_{s}\over g_{s}}(M-M^{*})\right]^{2}} + {\nu \over 2\pi^{2}}\int_{0}^{k_{F}} E^{*}(k') k'^{2} dk'+ \eta\Gamma_{\varepsilon}, 
\end{equation}
\begin{equation}
p'={1 \over 2}\left( { g_{v} \over m_{v}}\rho_{B}\right)^{2} - {1 \over 2}{\left[ {m_{s}\over g_{s}}(M-M^{*})\right]^{2}} +{1 \over 3}\left( {\nu \over 2\pi^{2}}\right) \int_{0}^{k_{F}} {k'^{4} \over E^{*}(k')} dk'+\eta\Gamma_{p}, 
\end{equation}
where
\begin{align}
\Gamma_{\varepsilon}={1 \over 2}{\nu \over (2\pi)^{3}}{1 \over \Omega} \int\int k'x'{\Omega_{\eta}\over E^{*}(k')}d\textbf{x}'d\textbf{k}',
\end{align}
\begin{align}
\Gamma_{p}=-{1 \over 12}{\nu \over (2\pi)^{3}}{1 \over \Omega} \int\int k'^{3}x'{\Omega_{\eta}\over [E^{*}(k')]^{3}} d\textbf{x}'d\textbf{k}',
\end{align}
where $E^{*}(k')=\sqrt{\textbf{k}^{'2}+M^{*2}}$ and $\Omega_{\eta}\equiv(\textbf{e}_{k'}\times\textbf{e}_{x'}).\textbf{e}_{\eta}$, where $\textbf{e}_{k'}$, etc. are the unit vectors in the direction of $\textbf{k}'$, etc.

Choosing $\eta=0$, we recover Eqs. (18) and (19). The relevant quantities for the NC NMEoS are given by:
\begin{equation}
\varepsilon_{NC}=\varepsilon_{C} + \eta\Gamma_{\varepsilon}, 
\end{equation}
\begin{equation}
p_{NC}=p_{C} +\eta\Gamma_{p}, 
\end{equation}
where  
\begin{align}
\Gamma_{\varepsilon}={1 \over 6}{\nu \over (2\pi)^{3}} \left( E_{F}^{*3}-3M^{*}E_{F}^{*}+2M^{*3}\right)\lambda_{\eta}
\end{align}
\begin{align}
\Gamma_{p}=-{1 \over 12}{\nu \over (2\pi)^{3}} \left[ \left( k_{F}^{2}-2M^{*2}\right) E_{F}^{*}-{2\over3}E_{F}^{*3}-M^{*2} {k_{F}^{2}\over E_{F}^{*}}+{8 \over 3} M^{*3}\right]\lambda_{\eta},
\end{align}
with $E_{F}^{*}=\sqrt{ k_{F}^{2}+M^{*2}} $ and $\lambda_{\eta}=\int \Omega_{\eta}x'd\textbf{x}'d\Omega_{k'}/\Omega$, where $d\Omega_{k'}$ is the solid angle element of $k'$. The factor $\lambda_{\eta}$ depends on the dimension of the box, here referred to as $^{\backprime\backprime}\textit{noncommutative geometry length}^{\prime\prime}$ ($\textit{NCGL}$). 

   The key quantity in the study of the NM properties is the nucleon \textit{effective mass}. The \textit{effective mass} in the NC case is given by:
\begin{align}
M^{*}=M-\left( {g_{s} \over m_{s}} \right)^{2} {\nu \over 4\pi^{2}} \int_{0}^{k_{F}} {M^{*}\over E^{*}(k')}dk'+\eta\Gamma_{M^{*}}, 
\end{align}
where
\begin{align}
\Gamma_{M^{*}}=\left( {g_{s} \over m_{s}} \right)^{2}{\nu \over (2\pi)^{3}} {M^{*} \over 4\Omega}\int\int {k'x'\over [E^{*}(k')]^{3}}d\textbf{x}'d\textbf{k}'.
\end{align}
As in Eqs. (33) and (34), the nucleon \textit{effective mass} in the NC geometry is given by:
\begin{align}
M^{*}_{NC}=M^{*}_{C}+\eta\Gamma_{M^{*}},
\end{align}
where
\begin{align}
\Gamma_{M^{*}}={1 \over 4}\left( {g_{s} \over m_{s}} \right)^{2}{\nu \over (2\pi)^{3}}M^{*2}\left( {E_{F}^{*}\over M^{*}}- {M^{*}\over E_{F}^{*}}-2\right)\lambda_{\eta}.
\end{align}
In the next section we present some numerical estimates for the NC effects.

\section{Results and Discussion}

\begin{figure}
\begin{center}

\includegraphics[scale=0.6]{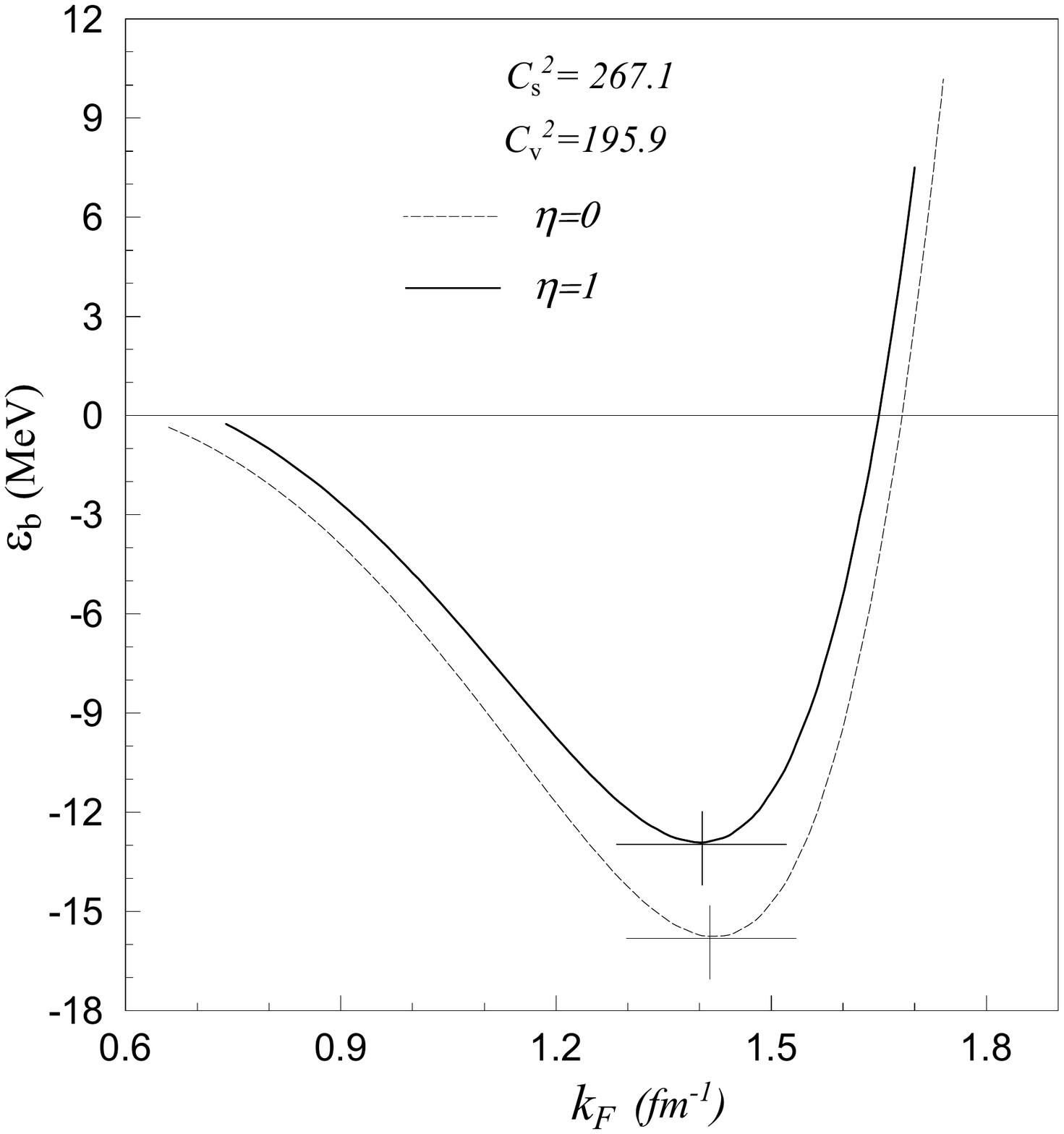}

\caption{Saturation curves of NM for two values $\eta$ (for $\lambda_{\eta}=1$) according to data of Table I.}\label{fig:binding}
\end{center}
\end{figure}

  We consider as input the saturation point values of NM in the usual QHD-I framework \cite{28}. We show in Table I, the standard values for the pertinent parameters, where $C_{s}^{2}=\left( {g_{s} \over m_{s}}M\right)^{2}$, $C_{v}^{2}=\left( {g_{v} \over m_{v}}M\right)^{2}$ and M is the average of proton and neutron masses.
\vspace{1.0 cm}
\begin{center}
{\footnotesize Table I Relevant values for the NM in the QHD-I framework \cite{28}.}
\\
\vspace{.5 cm}
\begin{tabular}{|c|c|c|c|c|}
\hline 
$M(MeV)$ & $ m_{v} (MeV)$ & $ m_{s} (MeV)$ & $C_{v}^{2}$ & $C_{s}^{2}$\\
\hline 
 938.93 & 782.6 & 550 & 195.5 & 267.1 \\
\hline 
\end{tabular} 

\end{center}
\vspace{1.0 cm}
  To investigate the effect of $\eta$ on the NM calculations, we consider the binding energy, $\varepsilon_{b}=\varepsilon/\rho-M$, of the symmetric NM in which $\rho_{p}=\rho_{n}=\rho_{B}/2$. Considering the trivial constraint $\rho_{B}=\rho_{p}+\rho_{n}$, we solve Eq. (29) and the self-consistent Eq. (37) with the assumption that $\lambda_{\eta}=1$. Fig.~\ref{fig:binding} shows the saturation curves of the symmetric NM for two different values of $\eta$ with respect to the coupling constants from Table I. For $\eta=0$ the saturation point takes place at $k_{F}=1.42 fm^{-1}$ with the value $\varepsilon_{b}=-15.75 MeV$ [28]. In the case of $\eta=1$, the saturation lies at $k_{F}=1.41 fm^{-1}$ with the value $\varepsilon_{b}=-12.86 MeV$. For $\eta\neq0$, although we cannot conclude that the saturation point is achieved, the qualitative behaviour of the curve is kept. 
  
\begin{figure}
\begin{center}

\includegraphics[scale=0.6]{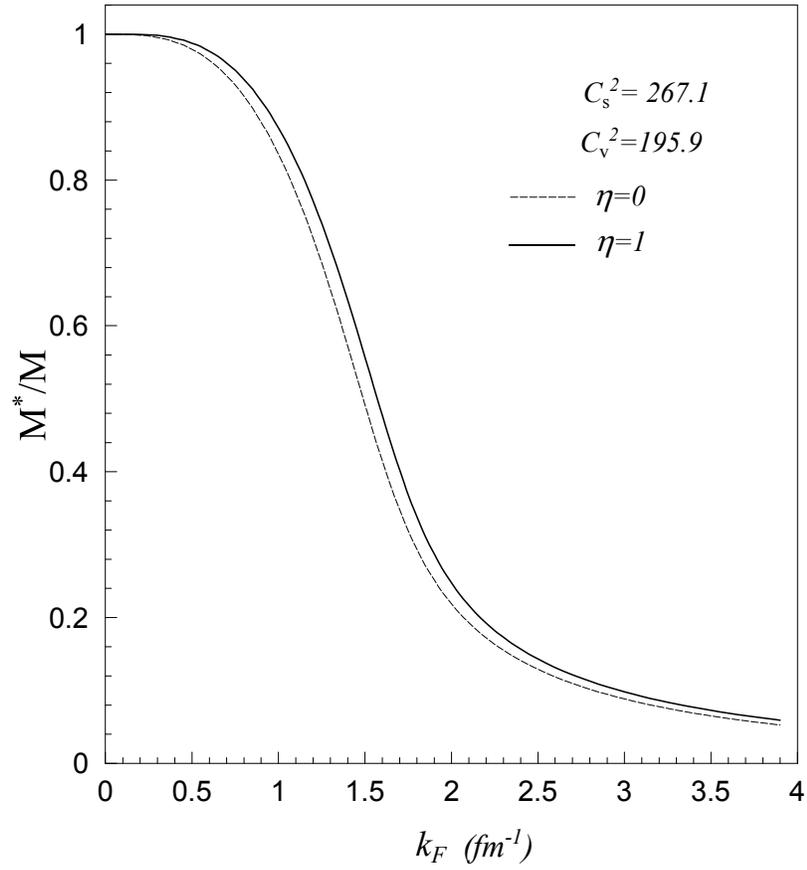}

\caption{The same as Fig. 1 for the \textit{effective mass}, $ M^{*}$.}\label{fig:effectivemass}
\end{center}
\end{figure}

  Fig.~\ref{fig:effectivemass} shows the behaviour of $M^{*}/M$ versus $ k_{F}$ for $\eta=0, 1$ for the parameters of Table I. It can be seen that $M^{*}$ is greater in the case of $\eta\neq0$ for $0.5 \lesssim k_{F} \lesssim 3.5 fm^{-1}$. As the \textit{effective mass} controls the stiffness of the NMEoS, the NC geometry softens the EoS of QHD-I model for the relevant values of $k_{F}$.
  
  To further investigate the effect of NC, we consider the following strategy. Applying the experimental saturation data of NM, $\varepsilon_{0b}=-15.86 MeV$ and $\rho_{0}=0.16 fm^{-3}$, and considering the trivial constraint on $\rho_{B}$, we calculate the coupling constants $g_{s}$ and $ g_{v}$, and $M^{*}$ through Eqs. (29), (30) and (37), for different values of $\eta_{NCGL}=\eta\lambda_{\eta}$. Table II shows the values of parameter $\eta_{NCGL}$ and the coupling constants which saturate the symmetric NM. The presented values of $\eta_{NCGL}$ have been calculated in the natural units, $\hbar=c=1$. As shown in Table II, the value of $M^{*}/M$, computed by solving the self-consistent equation under the empirical saturation conditions and suitable value of $\eta_{NCGL}$, lies in the acceptable interval, $0.7 \leq M^{*}/M \leq 0.8$ \cite{32}.
\vspace{.7 cm}
\begin{center}
{\footnotesize Table II The values of $\eta_{NCGL}$, $M^{*}/M$, $g_{s}\Phi_{0}$, $g_{v}V_{0}$, and the dimensionless coupling constants, which provide a fit for the experimental saturation data ($\varepsilon_{0b}=-15.86 MeV$ and $\rho_{0}=0.16 fm^{-3}$).}
\\
\vspace{.5 cm}
\begin{tabular}{|c|c|c|c|c|c|}
\hline 
$\eta_{NCGL}$ & $ M^{*}/M $ & $g_{s}\Phi_{0} (MeV)$& $g_{v}V_{0} (MeV)$ & $C_{s}^{2}$ & $C_{v}^{2}$\\
\hline 
$4.87\times10^{-3}$& 0.78 & 206.56 & 127.18 & 155.58 & 91.20\\
\hline 
$4.91\times10^{-3}$& 0.68 & 300.46 & 215.18 & 223.09 & 154.26\\
\hline 
\end{tabular} 

\end{center}
\vspace{.7 cm}

  The parameters in Table II show how the NMEoS changes as a function of the fundamental momentum scale, $\eta$. With respect to the magnitude of coupling constants from Tables I and II, the vector repulsive and scalar attractive parts of the symmetric NM in the case of NC are smaller than the usual case. This fact can be verified through comparing values of parameters $g_{s}\Phi_{0}$ and $g_{v}V_{0}$ for the NC case, Table II, and those of QHD-I where $g_{s}\Phi_{0}\simeq 400 MeV $ and $g_{v}V_{0}\simeq 330 MeV$ \cite{28}. The NC geometry reduces the interaction magnitude of propagating nucleons in constant scalar and vector fields. As can be seen, this reduction is approximately similar for both attractive and repulsive parts of the interaction.
   
   We can now compute the value of $\eta$ for the symmetric NM system. In order to do this, let us estimate the magnitude of the geometric parameter $\lambda_{\eta}=\int \Omega_{\eta}x'd\textbf{x}'d\Omega_{k'}/\Omega$, where $\Omega_{\eta}=(\textbf{e}_{k'}\times\textbf{e}_{x'}).\textbf{e}_{\eta}$. If we assume $ \textbf{e}_{k'}=\textbf{e}_{z'}$, $ \textbf{e}_{\eta}=\textbf{e}_{y'}$, and obtain $\Omega_{\eta}=1$; thus, $\lambda_{\eta}$ will be the double-integral over the solid angle element of $k'$ and the triple-integral over the magnitude of the position vector. Since the center of heavy nuclei is a typical example of NM, the root-mean-square radius ($R_{rms}$) of heavy nuclei can measure the volume of our box, $\Omega$. A straightforward calculation yields $\lambda_{\eta}\approx 70 fm $ for $R_{rms}\simeq 5.5 fm$. Therefore, an estimated value of $\eta$ is $6.96\times10^{-5}$ in the natural units, or $l_{\eta}\approx 0.014 MeV/c$. It should be noted that we use the value of $\eta$ corresponding to $M^{*}/M=0.78$. 

  We conclude, as a point of principle, that imposing the NC geometry in the QHD-I model, modifies the nuclear calculations and reduces the magnitude of nucleon-nucleon interaction selecting a suitable value of the NC geometry parameter, $\eta$. On the other hand, it leads to a softer NMEoS than that of the usual case. Since the NC geometry softens the NMEoS, we can expect, for instance, that NC in the mass-radius neutron stars calculations might lead to a smaller maximum mass than the usual case. Of course, the present scheme can be extended to the other nuclear systems, such as neutron matter and neutron stars, by including the effect of other particles such as leptons and hyperons. This might have relevant implications for the understanding of nuclear matter under astrophysical conditions (cf. Ref. [19]).

\pagebreak

\end{document}